\documentclass[aps,twocolumn,10pt,a4paper,showpacs]{revtex4}

\usepackage{graphics}
\usepackage{epsfig}

\begin{document}

\title{Application of Bryan's algorithm to the mobility spectrum analysis of semiconductor devices}

\author{D.~Chrastina}
\email{daniel@chrastina.net}
\altaffiliation{Current address: INFM and L-NESS Dipartimento di Fisica, Politecnico di Milano, Polo Regionale di Como, Via Anzani 52, I-22100 Como, Italy}
\affiliation{Department of Physics, University of Warwick, Coventry CV4 7AL, United Kingdom}
\author{J.~P.~Hague}
\altaffiliation{Current address: Department of Physics, University of Cincinnati, OH 45221-0011}
\affiliation{Department of Physics, University of Warwick, Coventry CV4 7AL, United Kingdom}
\author{D.~R.~Leadley}
\affiliation{Department of Physics, University of Warwick, Coventry CV4 7AL, United Kingdom}
\begin{abstract}
A powerful method for mobility spectrum analysis is presented, based
on Bryan's maximum entropy algorithm. The Bayesian analysis central to
Bryan's algorithm ensures that we avoid overfitting of data, resulting
in a physically reasonable solution. The algorithm is fast, and allows
the analysis of large quantities of data, removing the bias of data
selection inherent in all previous techniques. Existing mobility
spectrum analysis systems are reviewed, and the performance of the
Bryan's Algorithm Mobility Spectrum ({BAMS}) approach is demonstrated
using synthetic data sets. Analysis of experimental data is briefly
discussed. We find that {BAMS} performs well compared to existing
mobility spectrum methods.
\end{abstract}
\maketitle

\section{Introduction}

The characterization of semiconductor devices is essential for the
development of new materials which are interesting from both a physical
and technological point of view. Hall-effect measurements are very
frequently used for this characterization, especially when two-dimensional
systems are investigated.\cite{stradling}

The simplest ideal material has a single charge carrier with isotropic
energy bands and energy-independent scattering mechanisms. The resulting
longitudinal resistivity $\rho$ and Hall coefficient $R_H$ are constant
with respect to the application of magnetic field. Such a picture is
effective in a basic understanding of certain solid state phenomena.

In modern semiconductor devices, heterostructures are increasingly
employed where different semiconductor materials are grown epitaxially in
layers on a substrate. In this manner, structures can be designed in which
charge carriers are confined in a quantum well physically remote from the
dopant atoms they originate from; ionized impurity atoms can strongly
limit the mobility of charge carriers so this method of \emph{modulation
doping} can lead to greatly enhanced transport
properties.\cite{schaeffler1997,vkaenel2002,myronov2002}

However, the physics of magnetotransport in such heterostructures is
considerably more complicated if there are two or more distinct carrier
gases present in the material (for example, the intended carrier gas in
the quantum well plus the doped region) or the carrier gases feature a
spread of mobilities due to energy-dependent scattering mechanisms or
multiple subband occupancy. This leads to resistivities and Hall
coefficients which are dependent on the applied magnetic field, and to
extract the properties of such systems the simple single-carrier model is
not sufficient. To this end, various mobility spectrum techniques have
been employed.\cite{koschinksi1995,dziuba1997,vkaenel2002,myronov2002}

In a seminal paper, based on the work of McClure,\cite{mcclure} Beck
and Anderson\cite{beck} showed that the conductivity tensor depends on a
generalized conductivity function, $s(\mu)$, via the integral transforms,
\begin{equation}
\label{eq:sigxx}
\sigma_{XX}(B)=\int_{-\infty}^{\infty} \frac{s(\mu)}{1+\mu^2 B^2}\, d\mu
\end{equation}
and
\begin{equation}
\label{eq:sigxy}
\sigma_{XY}(B)=\int_{-\infty}^{\infty} \frac{\mu B s(\mu)}{1+\mu^2 B^2}\, d\mu
\end{equation}
where
\begin{equation}
\label{eq:s}
s(\mu) = n_s(\mu) e \mu
\end{equation}
and $n_s(\mu)$ represents the number (per unit area) of carriers with a
mobility $\mu$ averaged over the sample.

The Hall coefficient and resistivity are related to the elements of the
conductivity tensor in the standard way,
\begin{equation} \label{eq:rho}
\rho(B)=\rho_{XX}=
\frac{\sigma_{XX}}{\sigma_{XX}^2+\sigma_{XY}^2}
\end{equation}
\begin{equation} \label{eq:rh}
R_H(B)=-\frac{\rho_{XY}}{B}=-\frac{1}{B}\frac{\sigma_{XY}}{\sigma_{XX}^2+\sigma_{XY}^2}
\end{equation}

The function $s(\mu)$ appearing in Eqs.~(\ref{eq:sigxx})-(\ref{eq:s}) is
often known as the mobility spectrum. It gives the contribution to the
conductivity due to the density of carriers with mobility $\mu$. In this
investigation, it is assumed that $s(\mu)$ does not change with magnetic
field. At high fields and low temperatures this is not true because of the
formation of Landau levels: the density of states becomes a function of
magnetic field if $\hbar \omega_c > k_B T$ where $\omega_c$ is the
cyclotron frequency $eB/{m^{*}}$.\cite{coleridge:1989} Also, at very low
fields and low temperatures (such that $\mu B < \hbar / E_F \tau$ where
$E_F$ is the Fermi energy and $\tau$ is the momentum relaxation time)
quantum effects such as weak localization contribute to the
magnetoresistance.\cite{kearney:1992}

According to Eqns. (\ref{eq:sigxx}) and (\ref{eq:sigxy}) magnetoresistance
is always positive; negative magnetoresistance can only be analyzed by
assuming a magnetic-field dependent mobility or momentum relaxation
time.\cite{dziuba1999,dmitriev:2001}

Finding the magnetoresistance given a form for the mobility spectrum is
straightforward. However, performing the so-called inverse transform
problem (that is, finding the mobility spectrum from magnetoresistance
data) is non-trivial with a noisy, limited data set and is the subject of
much
investigation.\cite{beck,kim,kim1999,reginski,meyer,vurgaftman,kiatgamolchai,kiatgamolchai1}

We replace the traditional {MAXENT} algorithm used by Kiatgamolchai
\emph{et al.} \cite{kiatgamolchai,kiatgamolchai1} with Bryan's
algorithm.\cite{jarrell1996} This approach has two major advantages over
the previous technique: (1) A Bayesian analysis is performed on the
resulting spectrum, and (2) the search process is optimized by null space
decomposition of the kernel. The Bayesian aspect of Bryan's algorithm
ensures that we choose the most probable balance between the fit quality,
given by least squared minimization, and the information content of the
solution, given by the maximization of entropy. The null space
decomposition provides an exponential basis for the solution. The
exponential basis is particularly well suited to this type of problem
since it significantly reduces the number of dimensions in the search
space. It therefore allows reliable well-conditioned analysis of greater
quantities of data, avoiding the need for potentially biased data
selection. In this way, the results should be relatively free from common
artifacts such as mirror peaks, peaks of unnatural shape, or structure in
the spectrum at very low mobility values, provided that the artifacts are
not introduced by systematic measurement errors.\cite{achard}

We present this paper as follows. In Sec.~\ref{section:other}, early
approaches to this problem are reviewed, before considering the method of
maximum entropy in more detail. In Sec.~\ref{section:maxent}, the maximum
entropy method used in this paper is introduced, and technical details of
Bryan's algorithm are described. In Sec.~\ref{section:synthresults}, we
perform a case study on model data using our Bryan's Algorithm Mobility
Spectrum ({BAMS}) method. The usefulness of the mobility spectrum
technique for the analysis of real data is discussed in light of these
results, and a brief summary is given in Sec.~\ref{section:summary}.

\section{Background}
\label{section:other}

The approach of Beck and Anderson,\cite{beck} which first introduced the
concept of the mobility spectrum and has been developed into commercial
software, involves a multi-stage solution method (the numerical details of
which will not be explored). The number of data points used to create a
solution is equal to the number of distinct carriers gases within the
sample, plus one. The method either searches for an imposed number of
carrier gases or fits for a single carrier system, then a two-carrier
system, and so on. During each search for $n$ gases, the method tries
fitting to every available combination of $n+1$ data points from the
complete data set, only keeping the sets which lead to ``physical''
solutions. From the retained sets, the algorithm generates an ``envelope
function'' within which the mobility spectrum should be contained.

Once a set of carrier gases has been found, the validity of the solution
is confirmed by performing a least-squares fit on the original data with
respect to the carrier gas parameters. If the parameters emerge virtually
unchanged from the second stage, then the results of the original analysis
are considered to be reasonable.

If only a limited number of data points ($\sim10$) are available and the
system contains only two or three distinct (i.e. with mobilities which
differ significantly from each other) carrier gases, then this method is
quite efficient. However, with automated systems the data collected can
comprise thousands of values and the system simply cannot cope; choosing
points or creating averaged data by hand imposes artificial constraints on
the final solution. Also, this method does not respond well to errors in
the experimental data.\cite{kiatgamolchai}

Reference \onlinecite{kim} introduces and describes the Reduced
Conductivity Tensor method for extracting the carrier concentration and
mobility of each component of a multilayer semiconductor system. It is
assumed that the carrier gases are essentially degenerate and that the
effective mass is isotropic. These assumptions immediately remove the
possibility of gaining insight into energy-dependent scattering mechanisms
in non-degenerate systems, and also the possibility of analyzing for
example \emph{p}-type conduction in pure germanium in which the anisotropy
of the heavy hole band should be evident as ``harmonics'' in the mobility
spectrum.\cite{goldberg,mcclure,beer} Nevertheless, the method works well 
for up to three carrier gases.

In Ref. \onlinecite{reginski} it is assumed that the mobility for a 
gas subjected to mixed scattering can be approximated by the 
phenomenological expression:
\begin{equation}
\mu=\mu_0 x^\alpha \textrm{ where } x=\frac{E}{k_B T}
\end{equation}
and that the mobility spectrum of non-degenerate carriers in a spherical
band can be approximated by:
\begin{equation}
s(\mu)=S_0 e^{-x(\mu)} x(\mu)
\end{equation}
where $S_0$, $\mu_0$ and $\alpha$ are coefficients depending on the
density of states and on the parameters of the scattering mechanisms.
Then, the integrations in Eqs.~(\ref{eq:sigxx}) and (\ref{eq:sigxy}) are
converted to summations of discrete spectra for electrons and holes and an
iterative transformation procedure may be performed. The results are
decomposed into high and low-mobility carrier contributions. The
distinction between low-mobility carriers and high-mobility carriers is to
some extent arbitrary, however, it corresponds to the visible separation
between two different regions on the mobility spectrum. Results for the
high-mobility carrier gas seem much more satisfactory than results for the
low-mobility carrier gas in each case.

References \onlinecite{meyer} and \onlinecite{vurgaftman} describe the
\emph{quantitative mobility spectrum technique} ({QMSA}) and the
\emph{improved quantitative mobility spectrum technique} ({i-QMSA})
respectively. These are also iterative techniques but with no initial
assumption about the solution (although the Beck-Anderson\cite{beck}
approach is used to create a trial solution, and conductivity data is
extrapolated to higher magnetic field values to extend the available
mobility range). The {i-QMSA} method introduces a few extra tricks for
improving the fits whilst smoothing the spectra and making them more
physically reasonable. This method has also been successfully applied to
conduction in anisotropic bands by incorporating an explicit anisotropy
coefficient.\cite{vurgaftman:1999} In Ref.~\onlinecite{kiatgamolchai1} it
is shown that this method does not deal with error very well.

\section{Maximum Entropy}
\label{section:maxent}

A serious issue (regarding inverse transformation problems in general) is
that while a particular spectrum $s(\mu)$ may easily be transformed to
produce a magnetoconductivity $\sigma(B)$, the inverse problem is
ill-conditioned and the solution obtained for $s(\mu)$ by inverting the
kernel is extremely sensitive to small changes or errors in the
$\sigma(B)$ data. In particular, there is not necessarily a unique
$s(\mu)$ within the uncertainty bounds of the original data. For example,
since the integral transform has a blurring effect, then it is possible
for the data to have large local fluctuations while producing a smooth
$\sigma(B)$.

The technique of \emph{Maximum Entropy} has been employed to solve this
and other problems involving inverse integral transformation where the
result is a positive, additive
function.\cite{kiatgamolchai1,sivia,jarrell1996,chrastina,chu} Generally,
if two solutions (found by any means) are of equal merit (in terms of, for
example, their least-squares fits to the original results) then the
solution with the larger ``entropy'' is to be favoured, since it is
maximally non-committal with regard to missing (unmeasured) information in
the original data.\cite{kiatgamolchai1,jaynes} In other words, the
solution favoured by the maximum entropy method extracts the most
information out of the original data with the most reasonable assumptions
regarding information which is unavailable.

By discretising the mobility spectrum $\{s\}$, the entropy $S$ can be
defined as:\cite{gubernatis}
\begin{equation} \label{eq:ssj}
S\{s_j\} = \sum_{j=1}^N (p_j-m_j-p_j \ln p_j/m_j) 
\end{equation}
\begin{equation}
p_j = \frac{s_j}{\sigma_0}
\end{equation}
where $p_j$ is chosen so that the spectrum has a normalization of unity 
and $\sigma_0$ is the conductivity at zero magnetic field. The form of the
entropy imposes the condition that the mobility spectrum is non-negative,
which is physically realistic. If any prior information about the form of
the spectrum is available then this may be incorporated as a so-called
\emph{default model}~$\{m\}$.\cite{jarrell1996}

Fitting is now a matter of minimizing the function $Q$:\cite{jarrell1996}
\begin{equation} \label{eq:q}
Q=\chi^2 - \alpha S
\end{equation}
where $\chi^2 = \frac{1}{2}\sum_i
\left(\frac{\sigma(B_i)-\sigma_c(B_i)}{\delta_i}\right)^2$; $\sigma_c(B)$
is magnetoconductivity calculated from the fitting mobility spectrum,
$\delta$ is the error in the data such that a chi-squared which respects
the data has a value of approximately $M$ (the number of ``observations''
or data points) and $\alpha$ is a hyper-parameter which controls the
relative importance of the least-squares and entropic
constraints.\cite{jarrell1996}

There are various flavors of the maximum entropy method, which differ in
the way $\alpha$ is determined. In the case of historic maximum entropy,
different values of $\alpha$ are tried, until the minimization of $Q$
[Eq.~(\ref{eq:q})] gives $\chi^2=M$. In the case of classic maximum
entropy, the most probable value of $\alpha$ is found, given the data and
the default model.\cite{gubernatis,gallicchio} Alternatively, the method
described in Ref. \onlinecite{jaynes} treats the maximization of the
entropy as a natural starting-point for derivations of other results of
statistical mechanics.

The approach of Bryan's algorithm is to calculate spectra for a range of
$\alpha$ values and evaluate the probability that the solution is correct
given the data and the default model. A weighted average over $\alpha$ is
then taken.\cite{jarrell1996} However, this method may not work well if
$\alpha$ is very small. In the method developed in Ref. \onlinecite{hague}
a solution produced by Bryan's algorithm maximum entropy is used as the
default model for the next solution. This procedure is iterated until the
most probable solution corresponds to an $\alpha$ value well within the
range of applicability of Bryan's algorithm.

Since there are two integral transforms involved in calculating
$\sigma(B)$, then we work with\cite{kiatgamolchai1}
\begin{widetext}
\begin{equation}
\label{eq:sig}
\sigma(B) = \sigma_{XX}(B)+\sigma_{XY}(B)\nonumber =\frac{1}{2}\sigma_0 + 
\frac{1}{2} \int_{-\pi/2}^{+\pi/2} {s(\mu) [\mathrm{cos}({2 \theta}) + \mathrm{sin}({2 \theta})]\mathrm{d} \theta}
\end{equation}
\end{widetext}
where tan$(\theta) = \mu B$. This produces a fit to
$\sigma_{XX}(B)+\sigma_{XY}(B)$ rather than optimally to each component,
which could be seen as a drawback. Efforts towards the development of a
method which inverts each of the $\sigma_{XX}(B)$ and $\sigma_{XY}(B)$
tranforms in a semi-independent manner have not been successful so far.
However, since $\sigma_{XX}(B)$ is an even function and $\sigma_{XY}(B)$
is odd, then if both negative and positive magnetic field values are used
symmetrically, a fit to $\sigma_{XX}(B)+\sigma_{XY}(B)$ unambiguously fits
$\sigma_{XX}(B)$ and $\sigma_{XY}(B)$. This improves results, especially
for low mobility carrier gases, as will be seen in Sec.
\ref{section:synthresults}.

Since we have assumed discrete forms for $s(\mu)$ and $\sigma(B)$, then
Eq. (\ref{eq:sig}) can be rewritten $\sigma_i = K_{ij} s_j$ (with $M$ 
data points and $N$ mobility spectrum points). The mobility spectrum may 
be computed as:
\begin{equation} \label{eq:sigj}
s_j = K_{ij}^{-1} \sigma_i
\end{equation}

Since the magnetoconductivity $\sigma(B_i)$ is relatively insensitive to
the details of the mobility spectrum $s(\mu_j)$, the kernel contains a
large quantity of repeated information; many of the linear equations
described by the kernel are (almost) identical.\cite{press} In addition,
elements of the kernel matrix which are dominated by signal noise or
computational rounding errors are liable to make a huge contribition to
Eq. (\ref{eq:sigj}).

The dominant effects of data error in the problem can be investigated via
\emph{singular value decomposition} (SVD). The $M\times N$ kernel may
generally be rewritten as $K=UWV^T$ where $U$ is an $M\times N$
column-orthogonal matrix, $W$ an $N \times N$ diagonal matrix with
nonnegative values (the ``singular values'') and $V^T$ the transpose of an
orthogonal $N \times N$ matrix $V$. As a result of the orthogonalities,
the inverse of the kernel can simply be written as $K^{-1} =
VW^{-1}U^T$.\cite{press}

If one of the elements of $W$ is zero then the corresponding element of
$W^{-1}$ will be zero and the matrix is singular. However, small
(non-zero) elements resulting from inaccuracies invert to large values,
whereas they should become zeros. It is therefore justified to set values
of $W$ with magnitude smaller than the data error to zero.\cite{press} SVD
reduces the dimensionality of the space that must be searched for a
solution giving a massive increase in speed, and discards the effects of
noise which would corrupt the solution. It is therefore a very important
feature of the following analysis.

A simple, powerful and successful maximum-entropy method for finding a
mobility spectrum is described in Refs.~\onlinecite{kiatgamolchai} and
\onlinecite{kiatgamolchai1}, based on the method of Jaynes.\cite{jaynes}
The described procedure does not, however, use SVD to remove
overspecification and noise and so the calculation time increases with the
square of the product of the number of mobility and magnetic field points,
$(NM)^2$;\cite{kiatgamolchai} nor does it make explicit a hyper-parameter
[$\alpha$ in Eq. (\ref{eq:q})] for balancing fit quality against the
maximization of the entropy, or deal with error in the data in an obvious
way.  Also, there does not appear to be any way in which the likelihood of
the solution is checked and this leads to overfitting and the creation of
artifacts in the mobility spectrum when applied to real
data.\cite{myronov2002}

Minimising $Q$ leads to the following relation,
\begin{equation}
-\alpha\ln(\frac{s_i}{m_i})=\frac{1}{2}\sum_j K_{ij}^{T}\frac{\partial \chi^2}{\partial\sigma_j(B)}
\end{equation}

By introducing the SVD to this relation, it can be shown that the $s_i$
can be cast in the form of an exponential basis,
\begin{equation}
s_i=m_i\exp(\sum_{j=1}^s U^{(s)}_{ij}u_j)
\end{equation}
where the superscript $(s)$ represents matrices in the reduced space.

The solution may then be found in terms of the $s$-dimensional $u_t$ space 
rather than the $N_{output}$-dimensional $s_i$ space. The $u_t$ are found
iteratively by using a Newton method to obtain the increment in parameter
space $\delta u_j$,
\begin{equation}
\left((\alpha+\epsilon)\delta_{ij}+\sum_k M_{ik}D_{kj}\right)\delta u_j=-\alpha u_j-g_j
\end{equation}
where,
\begin{equation}
g_i=\frac{1}{2}w_{i}^{(s)}\sum_j V_{ij}^{(s)T} \frac{\partial \chi^2}{\partial\sigma_j(B)}
\end{equation}
and two new matrices,
\begin{equation}
\mathbf{M}=\frac{1}{2}\mathbf{W}^{(S)}[\mathbf{V}^{(S)}]^{T}\frac{\partial^2 \chi^2}{\partial\mathbf{\sigma_c}^{2}}\mathbf{V}^{(S)}\mathbf{W}^{(S)}
\end{equation}
\begin{equation}
\mathbf{D}=[\mathbf{U}^{(S)}]^{T}\mathrm{diag}[\mathbf{s}]\mathbf{U}^{(S)}
\end{equation}
are defined to speed up the algorithm: $\mathbf{M}$ can be calculated
once at the beginning of the algorithm, leaving the smaller computation of
$\mathbf{D}$ to the iterative process.\cite{touchette:1999}

Many solutions are found for different values of $\alpha$ and, by
considering the curvature of the parameter space in the vicinity of the
solution, the relative probability of the spectrum for a given $\alpha$
can be found. Performing the weighted average should then give the most
probable mobility spectrum.
\begin{equation}
\mathrm{P}[\alpha|\bf m ,G \rm ]\propto \prod_{i}\left(\frac{\alpha}{\alpha+\lambda_{i}}\right)^{1/2}\exp[-\chi^{2}/2+\alpha S]\mathrm{P}[\alpha]
\label{maxentprobalpha}
\end{equation}
$\lambda_{i}$ are eigenvalues of the matrix, $\Lambda_{i}$, which
describes the curvature of parameter space in the vicinity of the
solution,
\begin{equation}
\Lambda=\mathrm{diag}[\mathbf{s}^{1/2}]\mathbf{U}^{T}\mathbf{M} \mathbf{U}\mathrm{diag}[\mathbf{s}^{1/2}]
\end{equation}
and $\mathrm{P}[\alpha]$ (which must be rescalable) is normally given by
the Jeffries prior,\cite{jarrell1996}
\begin{equation}
P[\alpha]=1/\alpha
\end{equation}

In the BAMS method described below, the calculation time scales as
$N^{1.5}M^{0.6}$. The SVD scales as $N^{2}M$, and remains a few percent of
the calculation time.  For input data at 500 magnetic field values and an
output spectrum of 300 points (finding solutions at 1000 values of
$\alpha$) the first BAMS result is produced in roughly 3 minutes on a
1.7GHz \mbox{Intel Pentium 4} system running Linux, the first few seconds
of which are the SVD. Such a number of input points is realistic given
automated data-gathering systems; the time-limiting factor in such
experiments is often the rate at which the magnetic field can be swept.
All previous mobility spectrum analysis methods require that $\sim$~90~\%
of a typical data set were discarded by the user in order that the
algorithm would complete in a reasonable time or indeed work at all. The
SVD essentially performs this same task (of reducing the space that must
be searched for a solution) in a mathematically rigorous manner, removing
user bias and also saving time.

\section{Results on Synthetic Data}
\label{section:synthresults}

In order to characterize the BAMS solution method, synthetic data were
generated using Eqs.~(\ref{eq:sigxx})-(\ref{eq:rh}) with realistic
parameters and subject to varying degrees of noise. Carrier gases were
modelled as Gaussian peaks in $s(\mu)$ centered at the specified values.  
Since the conductivity of the carrier gas is given by the area under the
peak [see Eq.~(\ref{eq:sint})], the intended sheet carrier density and
peak width were also specified for each carrier gas. The \emph{height} of
a peak in $s(\mu)$ generally has no physical significance.

\begin{figure}[ht]
\begin{center}
\epsfig{file=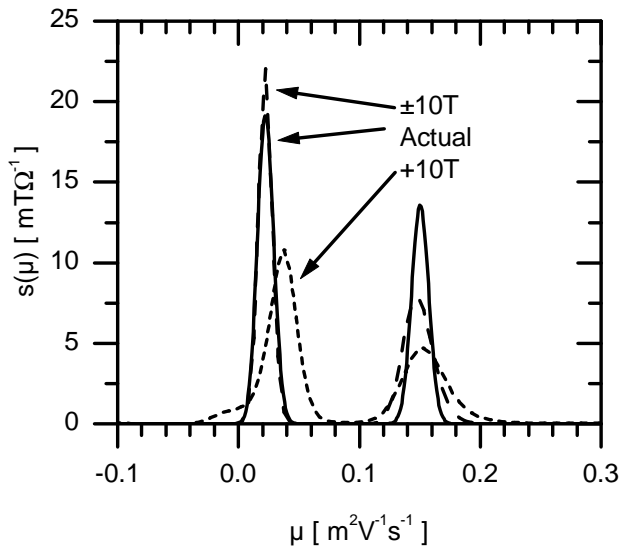, width=8.5cm}
\end{center}

\caption{Mobility spectra calculated on synthetic data featuring two
carrier gases of low mobility, with 0.01\% random data error. There were
200 output points in the mobility spectrum between
$\mu$~=~-0.5~m$^2$V$^{-1}$s$^{-1}$ and +0.5~m$^2$V$^{-1}$s$^{-1}$. It can
be seen that the result changes depending on whether only one field
direction (500 points) or both (1000 points) are used even though this
formally does not change the amount of information. Peak positions are
summarized in Table~\ref{tab:2lowsumm}.}

\label{fig:2plow}
\end{figure}

There were 1000 $(B,\sigma_{XX},\sigma_{XY})$ points from $-10$~T to 10~T
in each case, apart from Fig.~\ref{fig:2plow} where a dataset with 500
points from 0~T to 10~T is used for comparison.

\begin{figure}[ht]
\begin{center}
\epsfig{file=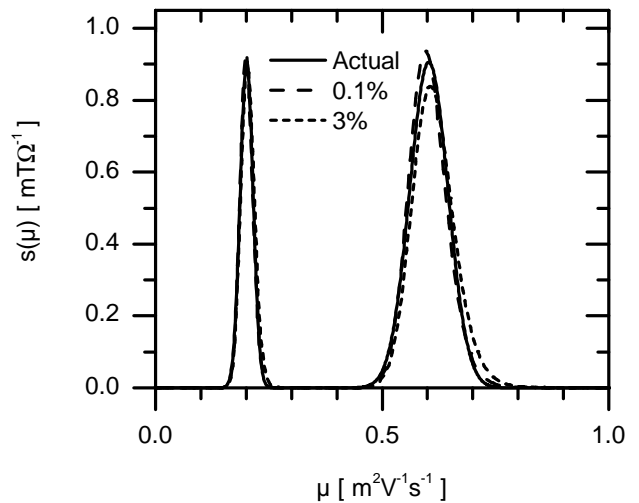, width=8.5cm}
\end{center}

\caption{Mobility spectra calculated on 1000-point synthetic data
featuring two carrier gases (the solid curve designated ``Actual'' is the
specified spectrum) with the specified level of random data error. There
were 1000 output points in the mobility spectrum between
$\mu$~=~-2.0~m$^2$V$^{-1}$s$^{-1}$ and +2.0~m$^2$V$^{-1}$s$^{-1}$ but only
the region of interest is shown. Even 3\% error in the data does not spoil
the result. Peak positions are summarized in Table~\ref{tab:2summ}.}

\label{fig:2phigh}
\end{figure}

The results shown in Fig. \ref{fig:2phigh} (for synthetic data created
from a spectrum of two Gaussian peaks with
$\mu_1$=0.2~m$^2$V$^{-1}$s$^{-1}$ and $n_1$=$1 \times 10^{11}$~cm$^{-2}$,
$\mu_2$=0.6~m$^2$V$^{-1}$s$^{-1}$ and $n_2$=$1 \times 10^{11}$~cm$^{-2}$;  
in this case the width of each peak is proportional to the peak mobility)
demonstrate how error in the data leads to very little broadening and
inaccuracy of peaks in the output spectrum. The iterative procedure where
one solution is used as the default model for the next [see Eq.
(\ref{eq:ssj})] was not necessary in this case. By integrating over each
peak (nominating a suitable $\mu_{peak}$ value) the number of carriers can
be found within each gas:
\begin{equation}
\label{eq:sint}
\int_{peak} s(\mu) \mathrm{d} \mu = n_s e \mu_{peak}
\end{equation}

Additionally, since the Hall sheet density is defined as $n_{Hall} = (e
R_H(B\to0))^{-1}$ then from Eqs.~(\ref{eq:sigxy}) and (\ref{eq:rh}):
\begin{equation} \label{eq:nhall}
n_{\mathrm{Hall}}=\frac{\sigma_0^2}{e \int{\mu s(\mu) \mathrm{d} \mu}}
\end{equation}
\begin{equation} \label{eq:muhall}
\mu_{\mathrm{Hall}} = \sigma_0 R_H(B\to0) =
\frac{1}{\sigma_0} \int{\mu s(\mu) \mathrm{d} \mu}
\end{equation}

The true (drift) carrier concentration is simply found by summing 
the carrier concentrations of each peak:
\begin{equation} \label{eq:ndrift}
n_{\mathrm{Drift}}= \sum_i | n_i |
\end{equation}
so the drift mobility is:
\begin{equation} \label{eq:mudrift}
\mu_{\mathrm{Drift}}= \frac{\sigma_0}{e \sum_i | n_i |}
\end{equation}

A summary of the results from Fig.~\ref{fig:2phigh} are shown in
Table~\ref{tab:2summ}; a simple algorithm was used to find local maxima in
the spectra. (Rather than fit a Gaussian to the spectrum to determine the
position of the peak, we simply find the highest point: this method is
therefore limited in its precision by the number of mobility points chosen
for the spectrum.) In addition to the peaks described, there were
generally a few much smaller peaks contributing in total to less than
0.5\% of the conductivity. (For clarity, not all of the results summarized
in Table~\ref{tab:2summ} are shown in Fig.~\ref{fig:2phigh}.)

\begin{table*}[ht]
\caption{Summary of results of BAMS analysis on the 2-peak synthetic data
shown in Fig.~\ref{fig:2phigh}. It can be seen that even up to a 3\% data
error the carrier gases are located accurately. Since the algorithm for
finding the peak position is very simple, the precision is dependent on
the number of mobility points.}
\begin{tabular}{rcccccc}

\hline \hline

Error & $\mu_1$ & $n_1$ & $\mu_2$ & $n_2$ & $\mu_H$ & $n_H$ \\
& [cm$^2$V$^{-1}$s$^{-1}$] & [$10^{11}$ cm$^{-2}$] & [cm$^2$V$^{-1}$s$^{-1}$] & [$10^{11}$ cm$^{-2}$] & [cm$^2$V$^{-1}$s$^{-1}$] & [$10^{11}$ cm$^{-2}$] \\

\hline

\emph{Specification} & \emph{2000} & \emph{1.00} & \emph{6000} & \emph{1.00} & \emph{5000} & \emph{1.60} \\

0.01\% & 1982 & 1.01 & 5946 & 1.00 & 5025 & 1.59 \\
0.03\% & 1982 & 1.01 & 5946 & 1.01 & 5026 & 1.59 \\
0.1\% & 1982 & 1.01 & 5946 & 1.01 & 5028 & 1.59 \\
0.3\% & 1982 & 1.01 & 5946 & 1.01 & 5025 & 1.59 \\
1\% & 2022 & 1.00 & 5986 & 1.00 & 5037 & 1.59 \\
3\% & 2022 & 1.04 & 6066 & 0.98 & 5057 & 1.59 \\

\hline \hline

\end{tabular}
\label{tab:2summ}
\end{table*}

Generally, when the peak mobility is overestimated the sheet density is
underestimated (and vice-versa) leading to a correct partial conductivity.
The Hall mobility compares well with the ``ideal'' value of
5000cm$^2$V$^{-1}$s$^{-1}$ at 1.6$\times 10^{11}$ cm$^{-2}$ (given by the
two-carrier Hall-effect equations).\cite{stradling} Even with 3\% error
in the data the results are acceptable.

With this software (as with that described in
Refs.~\onlinecite{kiatgamolchai} and \onlinecite{kiatgamolchai1}) it is
possible to find peaks in the mobility spectrum at mobilities smaller than
$1/B_{max}$, and this is shown in Fig. \ref{fig:2plow} and
Table~\ref{tab:2lowsumm}. The synthetic data feature two carrier gases
($\mu_1$=200~cm$^2$V$^{-1}$s$^{-1}$ and $n_1$=$1 \times
10^{13}$~cm$^{-2}$, $\mu_2$=1500~cm$^2$V$^{-1}$s$^{-1}$ and $n_2$=$1
\times 10^{12}$~cm$^{-2}$) with 0.01\% random data error.

\begin{table*}[ht]
\caption{Summary of results of BAMS analysis on the 2-peak synthetic data
shown in Fig.~\ref{fig:2plow}. If both field directions are used then the 
peaks are found more accurately. This effect is strongest for the low 
mobility peak.}
\begin{tabular}{rcccc}

\hline \hline

& $\mu_1$ & $n_1$ & $\mu_2$ & $n_2$\\
& [cm$^2$V$^{-1}$s$^{-1}$] & [$10^{12}$ cm$^{-2}$] & [cm$^2$V$^{-1}$s$^{-1}$] & [$10^{12}$ cm$^{-2}$] \\

\hline

\emph{Specification} & \emph{200} & \emph{10.0} & \emph{1500} & \emph{1.00}\\

-10T to +10T & 226 & 8.8 & 1482 & 1.0 \\
0T to +10T & 377 & 5.6 & 1533 & 0.9 \\

\hline \hline

\end{tabular}
\label{tab:2lowsumm}
\end{table*}

If only positive field values are used, the peaks are found at mobility
(density) values of 377~cm$^2$V$^{-1}$s$^{-1}$ ($5.6 \times
10^{12}$~cm$^{-2}$) and 1533~cm$^2$V$^{-1}$s$^{-1}$ ($9.0 \times
10^{11}$~cm$^{-2}$). The low mobility peak is not in exactly the correct
position, but the high mobility peak (which tends to be more important for
real systems) is located accurately.

However, if both positive and negative field values are used the peaks are
found to be much closer to the specified values, at mobility (density)
values of 226~cm$^2$V$^{-1}$s$^{-1}$ ($8.8 \times 10^{12}$~cm$^{-2}$) and
1482~cm$^2$V$^{-1}$s$^{-1}$ ($1.0 \times 10^{12}$~cm$^{-2}$). This is
because the use of positive and negative field values fits better to
$\sigma_{XX}$ and $\sigma_{XY}$ whilst the the use of one field direction
only fits to $\sigma_{XX} + \sigma_{XY}$.

\begin{table*}[ht]
\caption{Summary of results of BAMS analysis on the 3-peak synthetic data
shown in Fig.~\ref{fig:3phigh}. The effects of increasing iteration over
default models can be seen. The peak at lowest mobility is not found in
the correct position because of the limited mobility resolution and the
simple algorithm used to find the maximum.}
\begin{tabular}{rcccccc}

\hline \hline

& $\mu_1$ & $n_1$ & $\mu_2$ & $n_2$ & $\mu_3$ & $n_3$ \\
& [cm$^2$V$^{-1}$s$^{-1}$] & [$10^{12}$ cm$^{-2}$] & [cm$^2$V$^{-1}$s$^{-1}$] & [$10^{12}$ cm$^{-2}$] & [cm$^2$V$^{-1}$s$^{-1}$] & [$10^{12}$ cm$^{-2}$] \\

\hline

\emph{Specification} & \emph{270} & \emph{7.5} & \emph{1500} & \emph{1.00} & \emph{6000} & \emph{1.00} \\

1 iteration & 300 & 6.5 & 1421 & 1.08 & 5946 & 1.01 \\
3 iterations & 300 & 6.5 & 1461 & 1.06 & 5986 & 1.00 \\
10 iterations & 260 & 7.5 & 1461 & 1.07 & 6026 & 0.99 \\
30 iterations & 260 & 7.4 & 1461 & 1.08 & 6026 & 0.99 \\
100 iterations & 260 & 7.5 & 1502 & 1.05 & 6026 & 0.99 \\

\hline \hline
\end{tabular}
\label{tab:3summ}
\end{table*}

\begin{figure}[ht]
\begin{center}
\epsfig{file=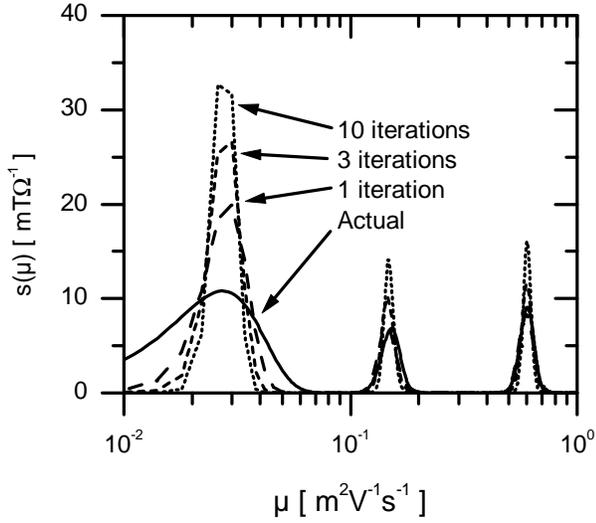, width=8.5cm}
\end{center}

\caption{Mobility spectra calculated on 1000-point synthetic data
featuring three carrier gases and 0.01\% data error, with increasing
iteration over the default model. There were 1000 output points in the
mobility spectrum (linearly spaced) between
$\mu$~=~-2.0~m$^2$V$^{-1}$s$^{-1}$ and +2.0~m$^2$V$^{-1}$s$^{-1}$ but only
the region of interest is shown. Peak positions are summarized in
Table~\ref{tab:3summ}. On this logarithmic scale it is possible to see
that the lowest mobility peak is only defined by a few spectrum points.}

\label{fig:3phigh}
\end{figure}

In Fig.~\ref{fig:3phigh}, which features three carrier gases, the
evolution of the spectrum with increasing iteration over default models is
shown. The positions of the peaks are summarized in Table~\ref{tab:3summ}.
With increasing iteration the peaks become too narrow, but the accuracy of
the determination of the peak position increases slightly. This
demonstrates that many iterations over successive default models is not
necessary; it also shows that this mobility spectrum analysis method may
not necessarily be relied upon the provide information regarding peaks
shapes or widths. Rather than converge on the true spectrum, with
increasing iteration over the default model the algorithm appears to
converge towards sharp narrow peaks.

\begin{figure}[ht]
\begin{center}
\epsfig{file=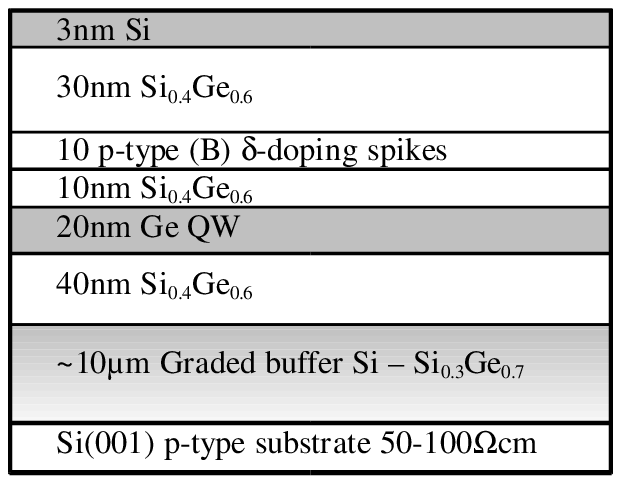, width=6.5cm}
\end{center}

\caption{The structure of device number 6016 grown by low-energy
plasma-enhanced chemical vapor deposition.\cite{vkaenel2002,kummer2002} It
features a strained 20~nm quantum well of Ge on a relaxed virtual
substrate of Si$_{0.3}$Ge$_{0.7}$. There are ten $\delta$ spikes of boron
10~nm above the channel. At 300~K there is conduction in both the quantum
well and the doped region so mobility spectrum analysis is required to
extract the properties of the 2-dimensional hole gas in the quantum well.}

\label{fig:6016}
\end{figure}

This algorithm has also been tested extensively on real data sets. For
example, magnetoresistance measurements on the device ``6016'' presented
in Ref.~\onlinecite{vkaenel2002} has been analyzed. The structure of this
device, grown using low-energy plasma-enhanced chemical vapor
deposition,\cite{kummer2002} is shown in Fig.~\ref{fig:6016}.

It features a strained 20~nm quantum well of pure Ge, on a relaxed virtual
substrate of Si$_{0.3}$Ge$_{0.7}$. The structure is doped with ten
$\delta$ spikes of boron 10~nm above the channel. This gives a sheet
density in the channel at 4.2~K of 6.2$\times$10$^{11}$cm$^{-2}$ with a
mobility of 87~000~cm$^2$V$^{-1}$s$^{-1}$.\cite{vkaenel2002}

At 300~K there is significant parallel conduction from the doped region,
giving a Hall mobility of 1790~cm$^{2}$V$^{-1}$s$^{-1}$ at a sheet density
of 1.8$\times$10$^{12}$cm$^{-2}$. We find a result for the channel
mobility at 300~K to be {2940~cm$^{2}$V$^{-1}$s$^{-1}$} with a sheet
density of 5.7$\times$10$^{11}$cm$^{-2}$ which agrees closely with the
quoted mobility value of {3000~cm$^{2}$V$^{-1}$s$^{-1}$}.

\section{Summary}
\label{section:summary}

A powerful mobility spectrum analysis method has been presented, based on
the application of Bryan's algorithm to the maximum entropy method. Our
method goes beyond a previous maximum entropy mobility spectrum
analysis\cite{kiatgamolchai,kiatgamolchai1} by using \emph{singular value
decomposition} (SVD) to reduce the dimensionality of the search space, and
consequently improve the conditioning of the problem. Thus,
overspecification and noise in the data are dealt with properly. This
means that a large data set of $100\sim1000$ points may be used directly
without a serious penalty in terms of calculation time. User bias with
regard to data point selection is therefore avoided. No other mobility
spectrum analysis method can deal with such large data sets without a
serious calculation time penalty.

No previous assumptions regarding the form of the result are required,
apart from the range over which to search for mobilities. Mobilities can
be accurately found which are smaller than $1/B_{max}$. Also, since the
algorithm works in a reduced space, there is no risk that peaks are lost
entirely if insufficient mobility resolution is specified.

This algorithm calculates mobility spectrum over a range of $\alpha$
values (where $\alpha$ controls the relative importance of $\chi^2$ versus
the entropy, that is, between a good fit and a physically reasonable
result)  and calculates the probability P[$\alpha$] of each spectrum.  
Then, the algorithm performs an average of the spectra weighted by their
calculated probabilities. The ensures that the solution is the best that
can be reasonably deduced from the data given
Eqs.~(\ref{eq:sigxx})-(\ref{eq:rh}). In this way, the results should be
relatively free from common artifacts such as mirror peaks, provided that
the artifacts are not introduced by systematic measurement
errors.\cite{achard}

However, there are a few caveats regarding mobility spectrum analysis in
general. It has been suggested\cite{kiatgamolchai} that the shapes of the
peaks in a mobility spectrum of a non-degenerate carrier gas could give
information regarding the energy-dependence of the momentum relaxation
time and therefore the dominant scattering mechanisms. However, the effect
on peak shape of different scattering mechanisms is very subtle and we
believe that it is unreasonable to expect any useful results from real
data. However, if {BAMS} analysis is performed on data from a range of
temperatures then the temperature dependence of the mobility can be
examined to give information about scattering mechanisms.

Also, whilst it should be possible to observe ``harmonics'' in a mobility
spectrum due to anisotropy of the effective mass\cite{mcclure} (for
example, in the heavy-hole band of germanium)\cite{goldberg} only 
Ref.~\onlinecite{vurgaftman:1999} seems to have explored this 
possibility.

Significant intersubband scattering can also invalidate mobility spectrum
analysis; in this case it is the intersubband scattering rates rather than
the transport scattering rates which must be considered.\cite{studenikin}

One final issue to consider is that the longitudinal magnetoresistance in
real data increases monotonically with field, whereas
Eqs.~(\ref{eq:sigxx})-(\ref{eq:rh}) predict that it should
saturate.\cite{goldberg} This is a serious issue when analyzing real high
magnetic field data, generic to all mobility spectrum techniques, and may
relate to the underlying assumption that $s(\mu)$ is independent of
magnetic field.\cite{argyres} Often the solution seems to be to assume
(possibly unreasonably) that there are mobilities which are much smaller
than $1/B_{max}$.\cite{contreras} However, using Bryan's algorithm to
balance the least squares fitting error with a physically reasonable
result ensures that this mobility spectrum method should minimize such
problems.

\begin{acknowledgments}
We would like to thank B. R\"{o}\ss ner and H. von K\"{a}nel for help and
support, and for making experimental data available. Parts of this work
were supported financially by the EPSRC and the INFM. Financial support
from GROWTH program ECOPRO No. GRD2-2000-30064 is gratefully acknowledged.
\end{acknowledgments}

\bibliography{bams}

\end{document}